\documentclass{ws-p8-50x6-00}

\begin{document}

\title{Systematics of Charged Particle Production in Heavy-Ion
Collisions with the PHOBOS Detector at RHIC}

\author{Peter A. Steinberg}

\address{
Brookhaven National Laboratory, Upton, NY 11973\\
E-mail: peter.steinberg@bnl.gov}

\author{\large PHOBOS collaboration}
\author { 
{
B.B.Back$^1$, M.D.Baker$^2$, 
D.S.Barton$^2$, R.R.Betts$^6$, R.Bindel$^7$,  
A.Budzanowski$^3$, W.Busza$^4$, A.Carroll$^2$,
J.Corbo$^2$, M.P.Decowski$^4$, 
E.Garcia$^6$, N.George$^1$, K.Gulbrandsen$^4$, 
S.Gushue$^2$, C.Halliwell$^6$, 
J.Hamblen$^8$, C.Henderson$^4$, D.Hicks$^2$, 
D.Hofman$^6$, R.S.Hollis$^6$, R.Ho\L y\'{N}ski$^3$, 
B.Holzman$^2$, A.Iordanova$^6$,
E.Johnson$^8$, J.Kane$^4$, J.Katzy$^{4,6}$, 
N.Khan$^8$, W.Kucewicz$^6$, P.Kulinich$^4$, C.M.Kuo$^5$,
W.T.Lin$^5$, S.Manly$^{8}$,  D.McLeod$^6$, J.Micha\L owski$^3$,
A.Mignerey$^7$, J.M\"ulmenst\"adt$^4$, R.Nouicer$^6$, 
A.Olszewski$^{3}$, R.Pak$^2$, I.C.Park$^8$, 
H.Pernegger$^4$, M.Rafelski$^2$, M.Rbeiz$^4$, C.Reed$^4$, L.P.Remsberg$^2$, 
M.Reuter$^6$, C.Roland$^4$, G.Roland$^4$, L.Rosenberg$^4$, 
J. Sagerer$^6$, P.Sarin$^4$, P.Sawicki$^3$, 
W.Skulski$^8$, 
S.G.Steadman$^4$, P.Steinberg$^2$,
G.S.F.Stephans$^4$,  M.Stodulski$^3$, A.Sukhanov$^2$, 
J.-L.Tang$^5$, R.Teng$^8$, A.Trzupek$^3$, 
C.Vale$^4$, G.J.van Nieuwenhuizen$^4$, 
R.Verdier$^4$, B.Wadsworth$^4$, F.L.H.Wolfs$^8$, B.Wosiek$^3$, 
K.Wo\'{Z}niak$^{2,3}$, 
A.H.Wuosmaa$^1$, B.Wys\L ouch$^4$
}
}
\address{
{
$^1$ Physics Division, Argonne National Laboratory, Argonne, IL 60439-4843,
$^2$ Chemistry and C-A Departments, Brookhaven National Laboratory, Upton, NY 11973-5000,
$^3$ Institute of Nuclear Physics, Krak\'{o}w, Poland,
$^4$ Laboratory for Nuclear Science, Massachusetts Institute of Technology, Cambridge, MA 02139-4307,
$^5$ Department of Physics, National Central University, Chung-Li, Taiwan,
$^6$ Department of Physics, University of Illinois at Chicago, Chicago, IL 60607-7059,
$^7$ Department of Chemistry, University of Maryland, College Park, MD 20742,
$^8$ Department of Physics and Astronomy, University of Rochester, Rochester, NY 14627,
}
}


\maketitle

\abstracts{
The multiplicity of charged particles produced in Au+Au collisions
as a function of energy, centrality, rapidity and azimuthal angle
has been measured with the PHOBOS detector at RHIC.
These results contribute to our understanding of
the initial state of heavy ion collisions and provide a means to
compare basic features of particle production in nuclear collisions
with more elementary systems.
}

\section{Introduction}
\vspace*{-.15cm}
Gold-gold collisions at RHIC provide the largest and potentially the 
densest partonic system available in the laboratory.  
The multiplicity of charged particles has been studied
as a function of center-of-mass energy ($\sqrt{s}$), 
centrality, rapidity and azimuthal angle
using the PHOBOS detector\cite{phobos-nim}, which covers
most of the available phase space.
These studies provide an understanding of the basic properties of the 
initial state, such as the effects of the collision geometry and 
parton saturation\cite{saturation-shapes}.
They also allow systematic comparisons of particle production in
nuclear collisions with more elementary systems like
proton-(anti)proton and proton-nucleus collisions. 

\vspace{-.3cm}
\section{Experimental Setup}

The PHOBOS detector\cite{phobos-nim}
consists primarily of silicon pad detectors mounted close
to the RHIC beampipe.  Charged particle 
multiplicity is measured in $4\pi$ using a combination of the ``Octagon''
detector which covers $|\eta| < 3$ (where $\eta = -\log \tan (\theta/2)$)
and a set of six ``Ring'' detectors
that cover $3 < |\eta| < 5.4$ with pads sized such that they cover 
approximately equal
bins in pseudorapidity. 
The vertex position is measured by two subdetectors, a high-precision vertex 
detector consisting of two silicon planes above and below the beampipe as well
as a 16-plane spectrometer covering $-1 < |\eta| < 2$.

To detect collisions and estimate centrality\cite{dndeta_cent} 
(here characterized as
the number of nucleons that interact inelastically), we use a combination
of two sets of 16 paddle counters and two
zero-degree calorimeters (ZDCs) located along the beam-line at distances of
3.28m and 18m from the interaction point in either direction,
respectively.

\section{Charged Particle Distributions}
\vspace*{-.15cm}
\begin{figure}[t]
\epsfxsize=15pc 
\epsfbox{prlfig3final.eps} 
\hspace{.5cm}
\begin{picture}(7,7)
\epsfxsize=10pc 
\put(0,0){\epsfbox{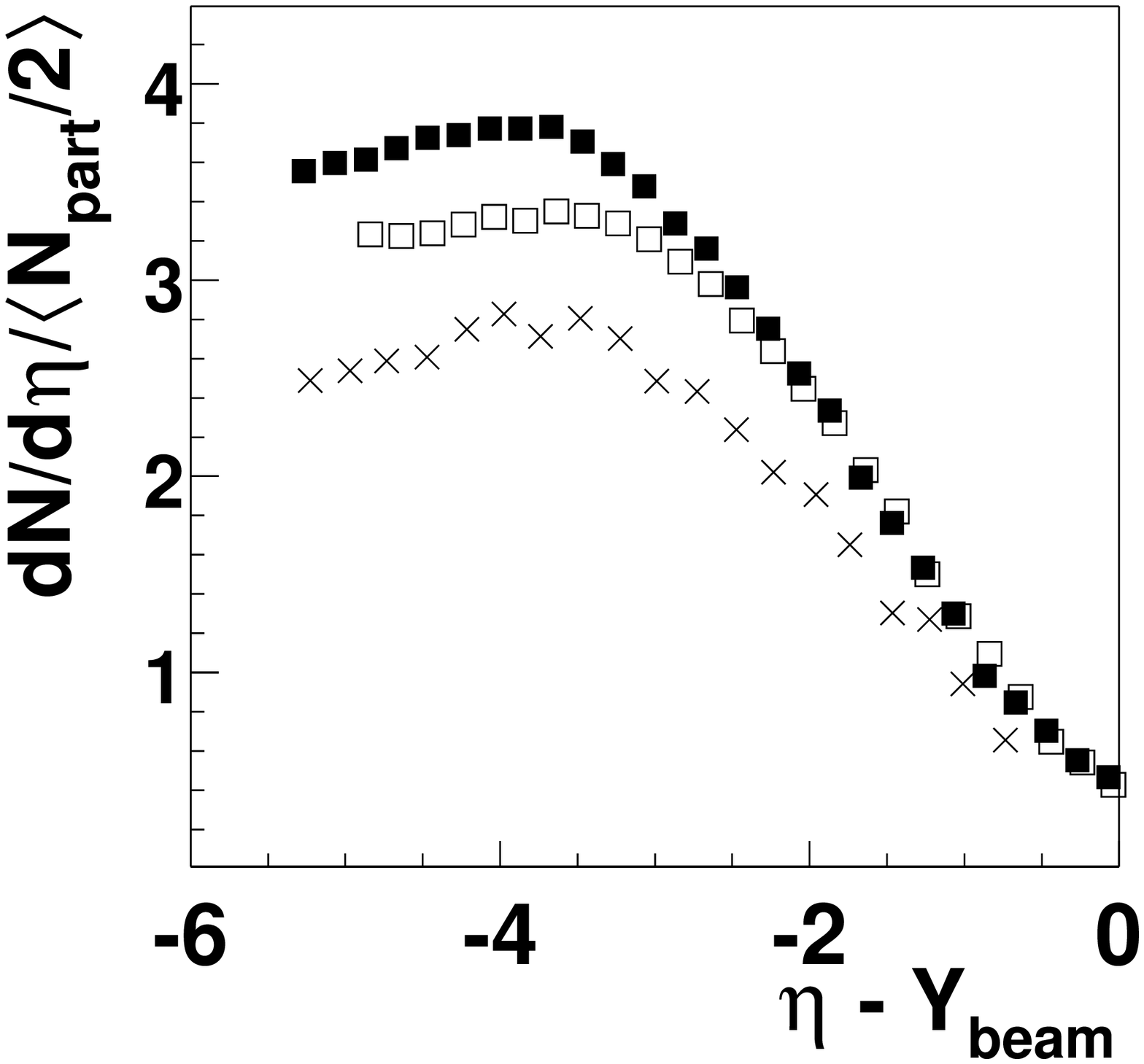}} 
\put(93,90){{\bf (c)}}
\end{picture}
\caption{
a.) Comparison of charged-particle pseudorapidity density for 
central (open circles) and peripheral events (solid points) at 
$\sqrt{s_{NN}}=130$ GeV
and b.)  Comparison of central data with
HIJING (solid line), AMPT (dashed line) and $p\overline{p}$ data (shaded band).  
c.) Pseudorapidity distribution for the most central events at $\sqrt{s_{NN}}=$ 130 (open squares) and 200 GeV (closed squares), shifted by the beam rapidity.
Limiting fragmentation is observed for $\eta - Y_{beam} > -2$.  
The crosses show data from UA5 for proton-antiproton collisions.  
Systematic errors are not shown, but are the same as in b.)
\label{dnde}
}
\vspace*{-.6cm}
\end{figure}

PHOBOS has produced the first measurements of $dN_{ch}/d\eta$ near mid-rapidity
at all RHIC energies to-date\cite{dndeta_200}, 
as well as detailed studies of the centrality dependence
of $dN_{ch}/d\eta$\cite{dndeta_cent}.
These measurements are important for
understanding the initial gluon density and 
are discussed elsewhere
in these proceedings\cite{itzhak}.

To measure $dN_{ch}/d\eta$ with the octagon and ring detectors,
we use two different techniques\cite{dndeta_shapes}.   
The first (``hit counting'') is 
based on a determination of the number of hits
in the silicon as a function of $\eta$.
We correct the raw hit multiplicity, as a function
of $\eta$, for occupancy (assuming Poisson statistics), 
geometrical acceptance, 
and background expected from Monte Carlo simulations.  
The second method is based on associating the energy deposition
in each pad with the charged particle multiplicity.

The data obtained at $\sqrt{s_{NN}}$ = 130 GeV shows several interesting features\cite{dndeta_shapes}.
As can be seen in Fig.\ref{dnde}a, 
while the multiplicity density 
scaled per participating nucleon pair
increases at mid-rapidity for more central collisions, 
it {\it decreases} outside of $|\eta| = 3-4$.  
Fig.\ref{dnde}b shows that the HIJING model\cite{hijing}, 
which includes no additional 
parton cascading, does not 
reproduce the data near $|\eta|=3-4$, unlike models (e.g. AMPT\cite{ampt}) 
which
include such effects.  Finally, by comparing central
events at $\sqrt{s_{NN}}$ = 130 and 200 GeV as a function of 
$\eta^\prime = \eta - Y_{beam}$ in Fig. \ref{dnde}b, we observe that the 
distributions obey ``limiting fragmentation'' for $ \eta^\prime> -2$
as seen in the UA5 $p\overline{p}$ data\cite{ua5}.
The UA5 distribution at $\sqrt{s} = 200$ GeV has
a shape similar to the PHOBOS 200 GeV data
but is 30-40\% lower at all pseudorapidities.

\vspace*{-0.4cm}
\section{Elliptic Flow}
\vspace*{-0.3cm}

\begin{figure}[t]
\epsfxsize=11pc
\epsfbox{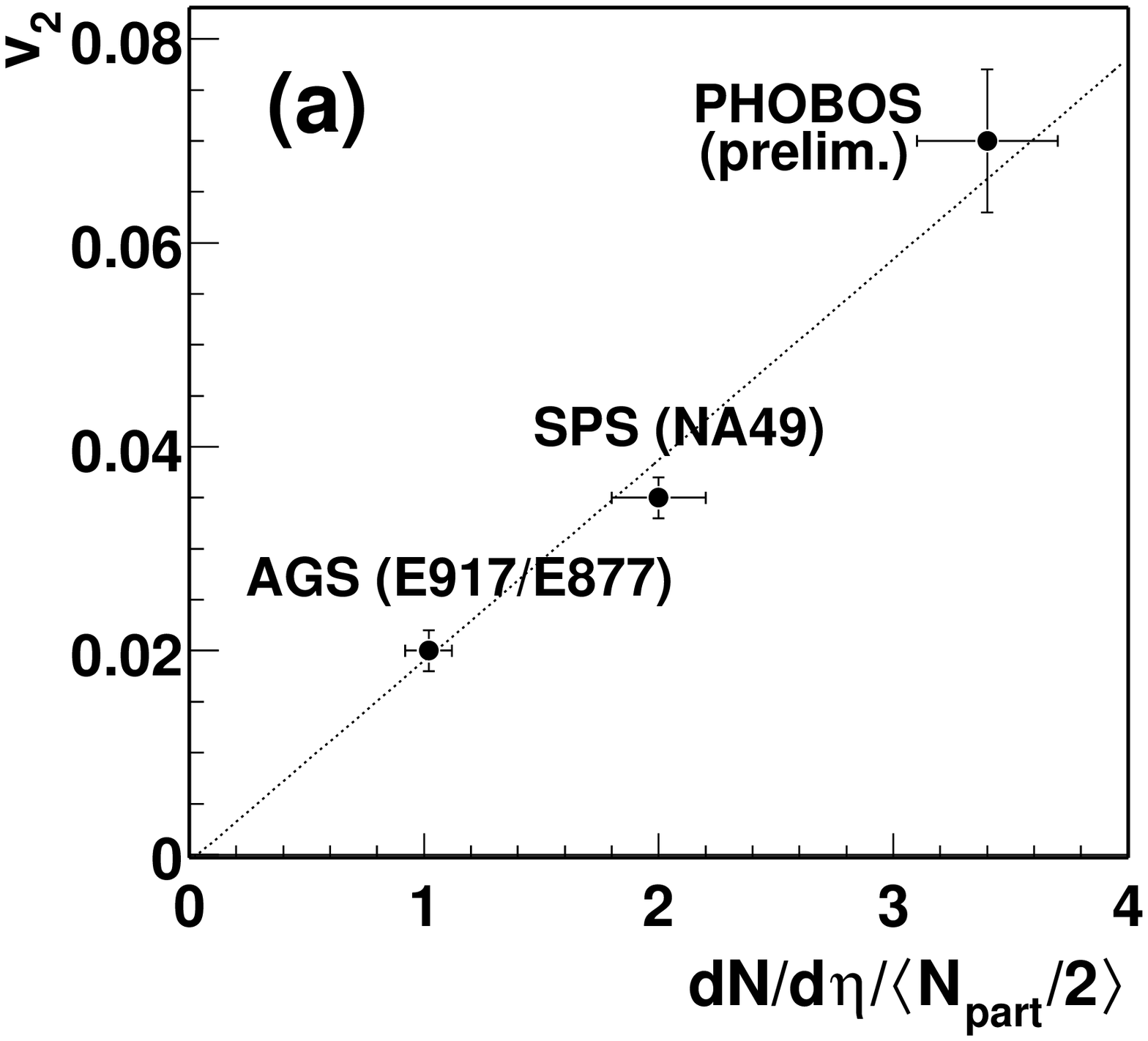}
\begin{picture}(10,7)
\epsfxsize=15pc
\put(0,0){\epsfbox{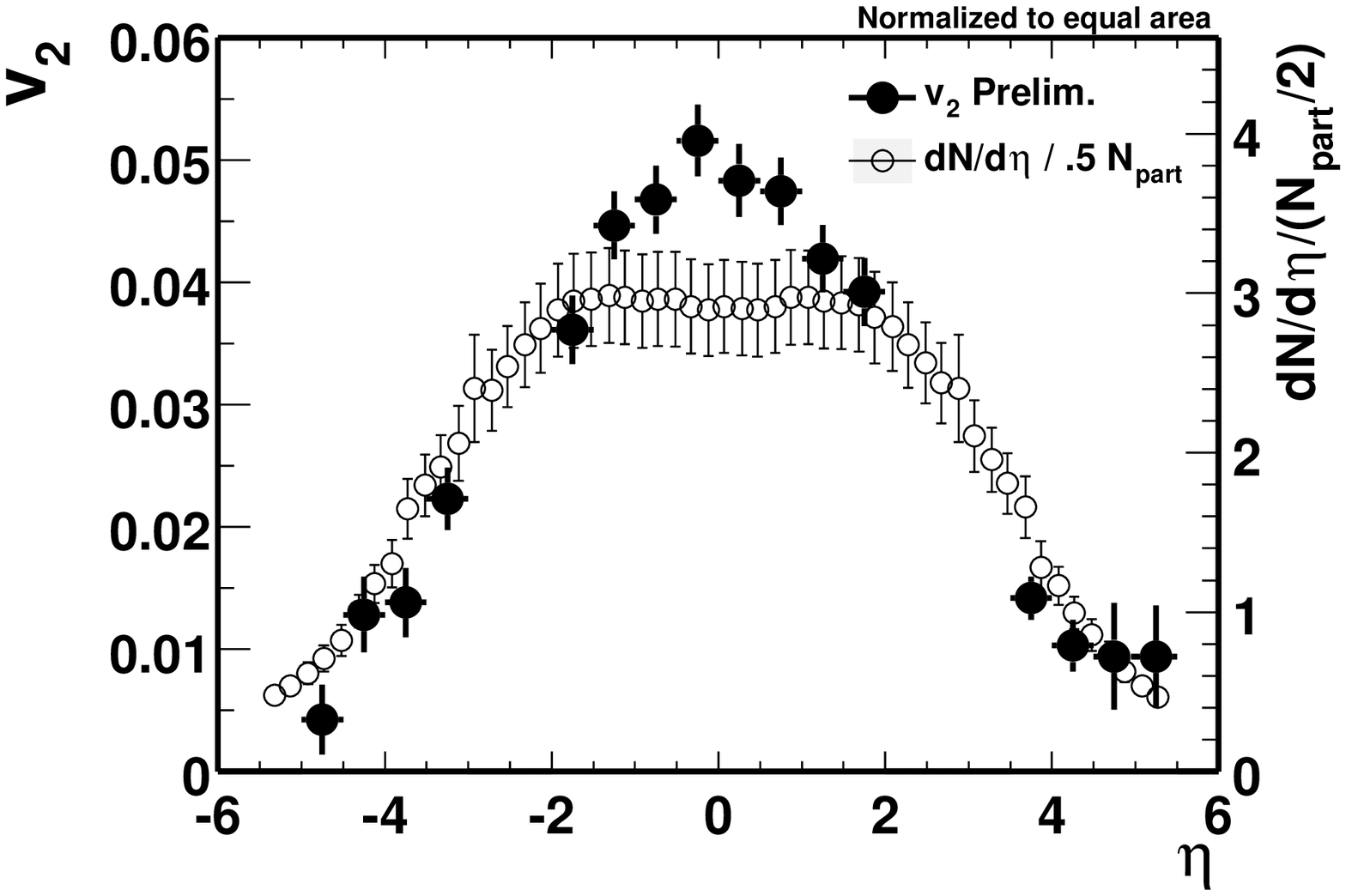}} 
\put(43,95){{\bf (b)}}
\end{picture}
\caption{a.) Data on $v_2$ vs. the multiplicity per participant pair for three different energies (AGS, SPS and RHIC).  b.) 
Preliminary data on $v_2$ vs. $\eta$ compared with a pseudorapidity density distribution of charged particles for a similar centrality, normalized to equal area.\label{flow}
}
\vspace*{-.5cm}
\end{figure}

To look for effects from initial state pressure gradients in peripheral
collisions, we have studied the second Fourier 
coefficient, $v_2$, of the azimuthal distribution 
relative to the reaction plane\cite{inkyu}.
We have measured $v_2$ vs. centrality
and found that it is at a maximum (5-6\%) in peripheral events and decreases
to 2-3\% in the 6\% most central events.  
This maximum value scales with energy in
a similar way to the charged-particle multiplicity, as shown in Fig. 
\ref{flow}a.  A preliminary analysis also indicates that $v_2$ vs. $\eta$ has a
shape approximately similar to $dN/d\eta$, as shown in Fig. \ref{flow}b.  
Both of these facts are consistent with the
flow signal reflecting the asymmetry in the particle density, 
with no additional contribution from hydrodynamic evolution.

\vspace*{-.2cm}
\section*{Conclusions}
\vspace*{-.15cm}
The PHOBOS collaboration has studied charged particle 
distributions as a function of
$\sqrt{s}$, centrality, rapidity, and azimuthal angle. 
The multiplicity studies provide information about the initial gluon density 
and provide a direct connection between high-energy $pp$ and $AA$ collisions. 
The elliptic flow results imply the presence of initial state pressure
gradients
but they are also broadly consistent with simple scaling with the 
rapidity density. 

\vspace*{-.5cm}
\section*{Acknowledgements}
\vspace*{-.2cm}
This work was partially supported by US DoE grants DE-AC02-98CH10886,
DE-FG02-93ER40802, DE-FC02-94ER40818, DE-FG02-94ER40865, DE-FG02-99ER41099, 
W-31-109-ENG-38 and
NSF grants 9603486, 9722606 and 0072204;
KBN of Poland grant 2 P03B
04916; and NSC of Taiwan contract NSC 89-2112-M-008-024.

\end{document}